\documentclass[aip,jcp]{revtex4-1}

\pdfoutput=1
\usepackage[utf8]{inputenc}
\usepackage[T1]{fontenc}
\usepackage[english]{babel}
\usepackage{amsmath}
\usepackage{graphicx}
\usepackage{float}

\usepackage{amsfonts}
\usepackage{amssymb}

\begin{document}

\title{Space- and time-dependent quantum dynamics of spatially indirect excitons in semiconductor heterostructures}
\author{Federico Grasselli}
\email{federico.grasselli@unimore.it}
\affiliation{Department of Physics, Informatics and Mathematics, University of Modena and Reggio Emilia, Italy}
\affiliation{CNR-NANO S3, Institute for Nanoscience, Via Campi 213/a, 41125 Modena, Italy}
\author{Andrea Bertoni}
\email{andrea.bertoni@nano.cnr.it}
\affiliation{CNR-NANO S3, Institute for Nanoscience, Via Campi 213/a, 41125 Modena, Italy}
\author{Guido Goldoni}
\email{guido.goldoni@unimore.it}
\affiliation{Department of Physics, Informatics and Mathematics, University of Modena and Reggio Emilia, Italy}
\affiliation{CNR-NANO S3, Institute for Nanoscience, Via Campi 213/a, 41125 Modena, Italy}

\date{\today}

\begin{abstract}
We study the unitary propagation of a two-particle one-dimensional Schr\"odinger equation by means of the Split-Step Fourier method, to study the coherent evolution of a spatially indirect exciton (IX) in semiconductor heterostructures. The mutual Coulomb interaction of the electron-hole pair and the electrostatic potentials generated by external gates and acting on the two particles separately are taken into account exactly in the two-particle dynamics. As relevant examples, step/downhill and barrier/well potential profiles are considered. The space- and time-dependent evolution during the scattering event as well as the asymptotic time behavior are analyzed. For typical parameters of GaAs-based devices the transmission or reflection of the pair turns out to be a complex two-particle process, due to comparable and competing Coulomb, electrostatic and kinetic energy scales. Depending on the intensity and anisotropy of the scattering potentials, the quantum evolution may result in excitation of the IX internal degrees of freedom, dissociation of the pair, or transmission in small periodic IX wavepackets due to dwelling of one particle in the barrier region. We discuss the occurrence of each process in the full parameter space of the scattering potentials and the relevance of our results for current excitronic technologies.
\end{abstract}

\maketitle

\section{Introduction}

Excitonic states are the fundamental below-gap resonances in semiconductors heterostructures, where stability and strength are strongly enhanced by quantum confinement.\cite{Bastard_BOOK88}
Spatially indirect excitons (IXs) are photo-excited correlated electron-hole pairs with the two charges localized in different segments of a nano-material. For example, IXs can be easily and selectively excited in semiconductor coupled quantum wells (CQWs), where an electric field applied \emph{perpendicular} to the plane of the quantum wells separates positive and negative charges in different layers.\cite{Butov_NatureA02,Sivalertporn_PRB12,High_PRL13}
Since optical recombination depends on the electron-hole overlap, this extends the intrinsic lifetime of excitons by orders of magnitude, from nanoseconds\cite{Colocci_EPL90} to microseconds,\cite{Butov_JPCM04,Gartner_APL06} which allows for the formation of new types of quantum condensates.\cite{Keldysh_JETP68,Butov_JPCM04,High_NATURE12,Alloing_EPL14}

In CQWs with an average inter-well separation $d$, IXs are two-dimensional quasi-particles in, say, the $(x,y)$ plane carrying a permanent electric dipole of order $-e d$ along the growth direction $z$. Beyond any change in the binding energy implied by the particle separation, an electric field $F_z$ shifts the energy of IXs by $U = -e d F_z$. Therefore, although electrically neutral, IXs couple to any gradient of $F_z(x,y)$ \emph{parallel} to the planes.\cite{Hagn_APL95,Gartner_APL06} Such energy landscapes $U(x,y)$ may be generated by imperfections and defects, but they can also be engineered by properly designed split gates, as in traditional nano-electronics, to modulate the perpendicular field in the planes. Within their extended lifetime IX can be moved over several micrometers,\cite{Gartner_APL06,High_Science11072008} while recombination can be controlled by switching off the perpendicular field,\cite{Winbow_NL07} broadening their application to an entirely new class of opto-electronic devices. In fact, in these devices data carried by photons can be stored in IXs, elaborated as in charge-based nano-electronics, and finally released back to photons by switching off the perpendicular electric field at the desired time, all processes being integrated in the same semiconductor segment.
Electronics based on excitons is called \emph{excitronics}. Since gating can be modulated at the GHz frequency, that is orders of magnitude shorter than the IXs intrinsic lifetime, excitronic systems set themselves as a natural bridge between optical data communication and nano-electronic devices.\cite{High_OL07,Kuznetsova_OL10,Andreakou_APL14}

Several excitronic functions have been demonstrated such as fast data storage,\cite{Winbow_NL07,Winbow_JAP08} acceleration with electrostatic ramps\cite{Gartner_APL06,Leonard_APL12} and interdigital devices,\cite{Winbow_PRL11} field effect transistors.\cite{High_Science11072008}
Most devices work in a diffusive regime, similar to traditional electronics. Therefore, IX are usually described theoretically as a classical gas of interacting dipoles.\cite{Winbow_PRL11,Wilkes_PRL12}
However, trapping of single IXs has been recently demonstrated,\cite{Schinner_PRL13} opening to the realization of quantum excitronic devices, similarly to single electron transistors (SET) in semiconductor quantum dots.\cite{Kastner_AP2000}

Clearly, for perspective devices it is important to determine the behavior of IXs in different regimes, in order to optimize the performance. For example, large dipoles may improve acceleration and operation rate, but ensuing smaller binding energies may cause dissociation and loss of the stored information. IXs are indeed polarizable quasi-particles, with both center-of-mass and internal dynamics. Since in typical systems applied electrostatic fields in the CQW planes and the mutual Coulomb interaction of the pair are both in the few meV range, the quantum dynamics of IX under the influence of accelerating potentials is a genuinely two-body process, and the internal quantum dynamics cannot be neglected in general. Although a fundamental problem, the quantum description of such two-body scattering process has been little discussed in literature.\cite{Arkhipov2003217}

In this paper we use a unitary propagation scheme to investigate theoretically the exact coherent dynamics of a single IX wavepacket describing scattering against different types of electrostatic potential landscapes, namely, steps/downhills and barriers/wells, using a minimal 1D model of a IX. In this sense, this is a two-particle generalization of the textbook description of one particle bouncing against potential steps and barriers. However, since the electron and the hole reside in different layers, in typical devices they are subject to different potentials. Therefore we have extensively investigated the parameter space separately for the two particles, analyzing the time-dependent dynamics during scattering as well as the long-time behavior of the scattered wavepacket. In addition to regions where IXs are reflected or transmitted almost as rigid objects, we identify experimentally relevant regimes where scattering is a genuine quantum two-body process. In these regimes, transmission or reflection of the wavepacket may be accompanied by excitation of the internal degrees of freedom, dissociation of the pair, or transmission in small periodic wavepackets due to dwelling of one particle in the barrier region.

In the next section (Sec.~\ref{sec:Model}) we set a 1D model of a IX and we briefly outline the numerical method used for propagation. In Sec.~\ref{sec:TDdynamics} we analyze the space- and time-dependent dynamics of IX wavepackets scattering against potential steps/downhills and barriers/wells. In Sec.~\ref{sec:scattering} we discuss the result of scattering at asymptotic times in the full parameter space of scattering potentials. Our results are discussed in Sec.~\ref{sec:conclusions} in connection with experimentally relevant regimes for current technologies. Details of the numerical method are given in Appendix~\ref{appendix_a}.
\section{Theoretical approach}
\label{sec:Model}

\subsection{The Hamiltonian}

To study the dynamics of IXs, we use a 1D model of an electron-hole pair propagating under the effect of the two-particle mutual Coulomb interaction and external electrostatic potentials coupling to the electron and hole separately. The two particles are tight to separated parallel channels at a distance $d$.
Accordingly, we consider the following electron-hole Hamiltonian
\begin{eqnarray}
\hat{H}=\hat{T}+\hat{U}_C+\hat{U}_{ext}, \label{ham_ecc_lib}
\end{eqnarray}
where
\begin{eqnarray}
\hat{T} & = & \frac{\hat{p}_e^2}{2 m_e} + \frac{\hat{p}_h^2}{2 m_h}, \\
\hat{U}_C & = &  - \dfrac{1}{\epsilon_r} \dfrac{e^2/ (4 \pi \epsilon_0)}{\sqrt{(\hat{x}_e - \hat{x}_h)^2 + d^2}} ,\\
\hat{U}_{ext} & = &  U_e(\hat{x}_e,t) + U_h(\hat{x}_h,t). \label{total_pot}
\end{eqnarray}
Here $\hat{p}_{e(h)}, \hat{x}_{e(h)}, m_{e(h)}$ are the 1D linear momentum operator, the coordinate operator and the effective mass of the electron (hole), respectively. $\hat{U}_C$ is the electron-hole Coulomb interaction in a medium of relative dielectric constant $\epsilon_r$.
$\hat{U}_{e(h)}(\hat{x}_{e(h)},t)$ represents the one-body electrostatic potential which couples to the electron (hole) coordinate $\hat{x}_{e (h)}$. Note that in our model the internal dynamics of the electron-hole pair is described in full, i.e., we do not assume any rigid exciton approximation,\cite{Zimmermann_PAC97,Hohenester_APL04}  nor the two particles are bound by construction: scattering against an external potentials may indeed result in excitation or dissociation of the pair. On the other hand, we neglect any lateral extension of the quantum states. Energy gaps induced by lateral confinement in a quantum well or wire (tens of meV) far exceed, and therefore decouple from, the small binding energy of IXs (few meV). Hence, the main effect of a lateral extension $L$ is to regularize the Coulomb interaction at short distances to $1/L$,\cite{Esser_PSSB01} giving just a small renormalization of the parameter $d$. Finally, note that in $U_{e(h)}(\hat{x}_{e(h)},t)$ we indicated a possible explicit dependence on time $t$. Indeed, the numerical methods which we discuss below also apply to the case of time-dependent potentials, although in this work we limit ourselves to stationary ones.

\begin{figure}[!h]
\includegraphics[scale=.6]{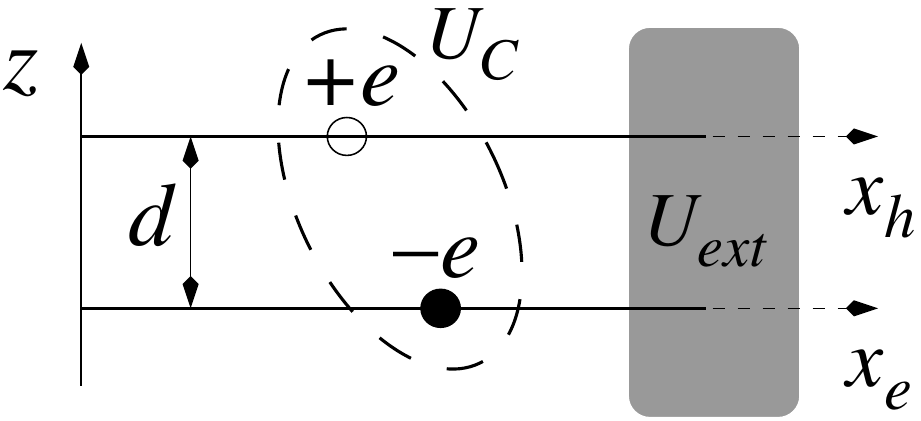}
\caption{Schematics of the 1D model assumed in the present work.}
\label{eccit1D}
\end{figure}

Below we shall discuss simulations performed with two paradigmatic classes of electrostatic potentials, namely, potential steps/downhills and potential barriers/wells. In Sec.\ref{sec:conclusions} we discuss the experimental relevance of these idealized potential profiles. A smooth potential step in the electron (hole) coordinate is described as a Fermi-like profile
\begin{equation}
U_{e(h)}(x_{e(h)}) = U^{e(h)}_0  \frac{1}{1 + e^{-(x_{e(h)}-b)/a}}. \label{pot_step}
\end{equation}
$x_{e(h)}=b$ is the position of the step, $U_{e(h)}(b)= \frac{1}{2} U^{e(h)}_0$. For a particle hitting from the left, $U^{e(h)}_0 >0 $ represents a repulsive potential energy step, while $U^{e(h)}_0 <0 $ represents an attractive potential energy downhill.  $a$ is a smoothness parameter. Using a smooth potential avoids inaccuracies in numerical work, particularly when numerical derivatives need to be computed, and gives a better model of realistic potential modulations.\footnote{This is not important in our simulation due to the diagonal representation of the kinetic operator, but it may be critical for {\it a posteriori} calculation of average quantities, like the kinetic energy.}
For simplicity, we assume common values $a$ and $b$ for electrons and holes.

A potential barrier is described by a double Fermi-like profile
\begin{equation}
\begin{split}
&U_{e(h)}(x_{e(h)}) =  U^{e(h)}_0 \times \\
&\times \left[ \frac{1}{1 + e^{-(x_{e(h)}-b_1)/a}} + \frac{1}{1 + e^{+(x_{e(h)}-b_2)/a}} -1  \right] \label{pot_barr}
\end{split}
\end{equation}
Here, $b_1, b_2$ are the mid-potential positions for the rising and descending parts of the barrier, respectively. $ U^{e(h)}_0>0$ represents a potential barrier, while $ U^{e(h)}_0<0$ represents  a potential well.

In the absence of any external potential, namely for a free IX, it is natural to adopt the center of mass (CM) and relative coordinate system
\begin{equation}
\left\{
\begin{array}{l}
X = \frac{m_e x_e + m_h x_h}{m_e + m_h} \\
x = x_e - x_h
\end{array}
\right.
\label{xr2xe}
\end{equation}
since the free Hamiltonian $\hat{H}_0 = \hat{T}+\hat{U}_C$ splits into
$\hat{H}_0 = \hat{H}_{CM} + \hat{H}_r$
where
\begin{eqnarray}
\begin{split}
\hat{H}_{CM} = \dfrac{\hat{P}^2}{2M} ,
\end{split}
\end{eqnarray}
and
\begin{eqnarray}
\begin{split}
\hat{H}_r = \dfrac{\hat{p}^2}{2m} - \dfrac{1}{\epsilon_r} \dfrac{e^2/ (4 \pi \epsilon_0)}{\sqrt{\hat{x}^2 + d^2}}.
\end{split} \label{rel_ham}
\end{eqnarray}
Here $P,p$ are the CM and relative momenta, respectively, while $M~\equiv~m_e+m_h$ and $m \equiv m_e m_h /M$ are the total and the reduced masses. The free IX wave function can be factorized as
\begin{eqnarray}
\Psi(X,x,t) = \Phi_K(X)\phi_n(x) e^{-iEt/\hbar},
\label{eq:factorization}
\end{eqnarray}
where the functions $\Phi_K(X)$ and $\phi_n(x)$ are eigenfunctions of $\hat{H}_{CM}$ and $\hat{H}_r$ with eigenenergies $E_{CM,K}$ and $E_{r,n}$, respectively. The CM component is labelled by the CM wavevector $K$ and the relative component by the integer quantum number $n$. The zero of energy is set at the energy gap of the quantum well material, i.e., the energy for generating the two free particles at rest. Therefore $E=E_{CM,K} + E_{r,n}$ is the total energy of the IX.

\begin{figure}
\includegraphics[scale=.6]{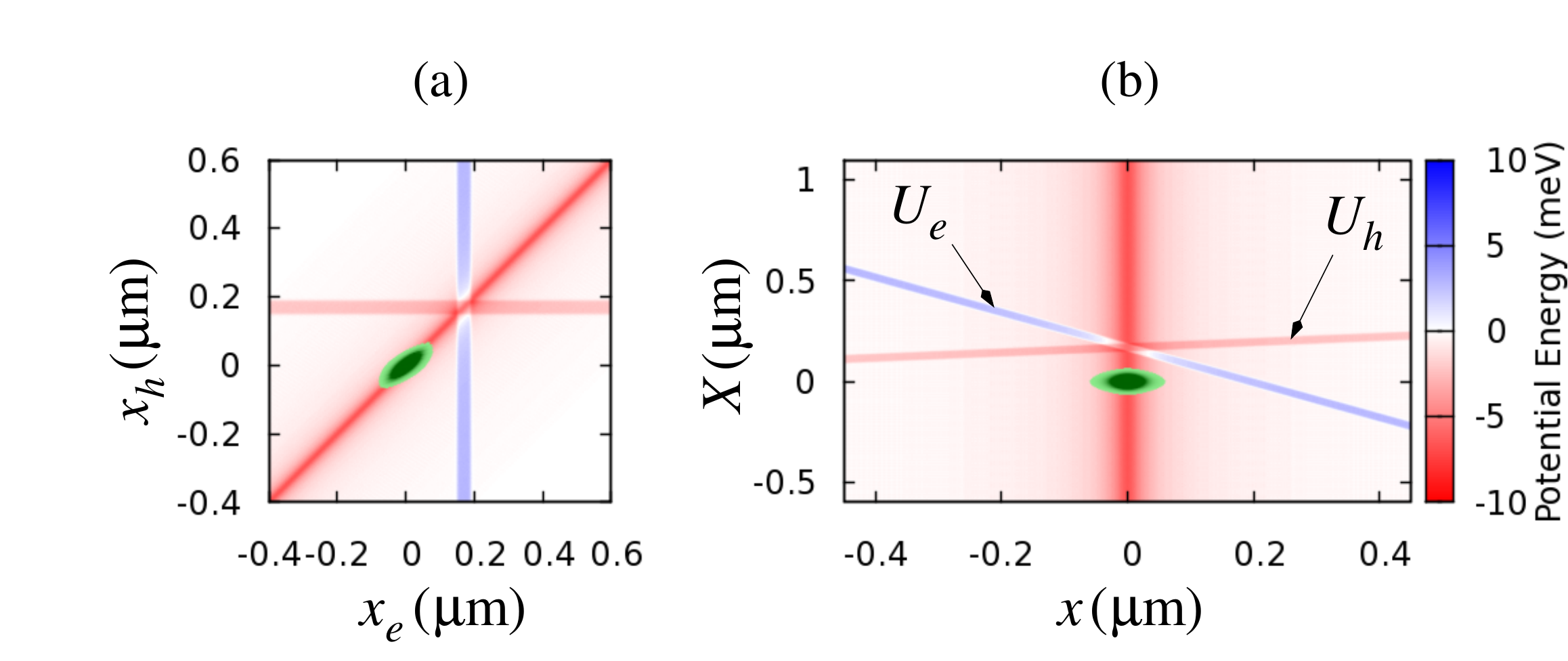}
\caption{Colormap of the potential landscape for an electron-hole pair in the presence of a barrier potential for the electron and a well for the hole with $a=2\,\mathrm{nm}, b_1=150\,\mathrm{nm} , b_2 =190\,\mathrm{nm}, U_0^e =+3.0\,\mathrm{meV}, U_0^h = -2.0\,\mathrm{meV} $ as in Eq.~(\ref{pot_barr}). (a): $(x_e,x_h)$ representation. (b): $(X,x)$ representation. Also indicated is the initial wavepacket, in arbitrary units, used in the simulations as described in Sec~\ref{sec:TDdynamics}. The arbitrary origin is set at the center of the initial IX wavepacket.}
\label{fig_init_state}
\end{figure}

In the presence of an external potential there is no general solution in closed form, since coordinates $(X,x)$ are coupled by $\hat{U}_{ext}$. On the other hand, had we used the $(x_e,x_h)$ coordinates which are separable in $\hat{U}_{ext}$, they would have been coupled by $\hat{U}_C$. In principle, the coherent propagation of a IX wavepacket can be carried out in both coordinate frames. However, as we aim at studying scattering events, we find the $(X,x)$ representation more convenient to analyze the result at asymptotic times.
Using the $(X,x)$ set of coordinates turns out to be also beneficial from the numerical point of view. In fact, electrons and holes are usually in the same region of space, due to Coulomb attraction. As we use a real space reprensentation, it is sufficient to describe small regions in $r$, that is $r_e \sim r_h$, rather then the full $(x_e,x_h)$ space.

Comparing the potential $\hat{U}_C+\hat{U}_e+\hat{U}_h$ in the two different coordinate systems may be useful to interpret the results of the following sections. In Fig.~\ref{fig_init_state} we show the total potential in the case of a narrow barrier for the electron and a well for the hole, together with the square modulus of the initial IX wave function. The broad negative strip represents the Coulomb potential, while the narrow straight lines are the electron and hole single particle potentials, $\hat{U}_e$ and $\hat{U}_h$. In the $(x_e, x_h)$ coordinate frame [Fig.~\ref{fig_init_state}(a)] $\hat{U}_e$ and $\hat{U}_h$ are represented by straight stripes parallel to the axes, while the Coulomb interaction couples the two coordinates, being maximum along the diagonal $x_e = x_h$. On the contrary, in the $(X,x)$ coordinate frame [Fig.~\ref{fig_init_state}(b)] the Coulomb interaction only depends on the relative coordinate $x$, while the external potentials $\hat{U}_e$ and $\hat{U}_h$ couple the $(X,x)$ coordinates and therefore are represented by the two oblique stripes, the one with negative (positive) slope acting on the electron (hole) alone. Note the difference in the absolute values of these slopes, ensuing from the very different effective masses.

\subsection{Numerical propagation}

The quantum propagation of the IX is performed on a discrete homogeneous time grid, evolving the state between two consecutive times $t$ and $t+\Delta_t$ by applying the evolution operator $\hat{\mathcal{U}}(t+\Delta_t,t)$ to the wave function:
\begin{equation}
\Psi(\mathbf{x},t+\Delta_t)=\hat{\mathcal{U}}(t+\Delta_t,t)\Psi(\mathbf{x},t). \label{tdprop}
\end{equation}
The evolution operator is applied by using the Split-Step Fourier (SSF) method, which is numerically exact for vanishing $\Delta_t$.\cite{Leforestier_JCP91,Castro_JCP04} Although this method is commonly exploited for the propagation of a single-particle wavepacket in a multi-dimensional physical domain, in Eq.~(\ref{tdprop}) we set $\mathbf{x}=(X,x)$, i.e, we interpret the two degrees of freedom of the system as the CM and relative coordinates of the electron-hole couple.

The SSF method relies on the Suzuki-Trotter expansion\cite{Suzuki_PJAB93} to factorize the evolution operator as a product between exponential operators, each of which is diagonal either in direct or in Fourier space (details are given in Appendix~\ref{appendix_a}). For this reason, this numerical method relies on a massive use of Fast Fourier Transformation (FFT), this resulting into a high efficiency and a relatively low computational cost in comparison with other methods. The use of FFT also implies periodic boundary conditions in $x$ and $X$. However, the simulation domain is chosen sufficiently large that the wave function does not reach the boundaries in our analyses.
Even if in this work we have been only concerned about time-independent potentials, the SSF method is also suitable for time-dependent ones.

\subsection{Initial state \label{init_state}}

The time-dependent analysis starts from the choice of a realistic initial state.
In a photo-generated exciton gas, the interaction with phonons and other scattering mechanisms allows for exciton relaxation to the ground state $\phi_0(x)$ within a few fs, i.e., on a much shorter timescale than photo-recombination lifetime. Therefore, at $t=0$ we consider IXs to be in their ground state $\phi_0(x)$ and we choose
\begin{equation}
\Psi(X,x,t=0) \equiv \chi(X) \phi_0(x).
\end{equation}
where $\phi_n(x)$, already defined in Sec.~\ref{sec:Model}, are computed numerically via finite-difference diagonalization, and
where $\chi(X)$ is a minimum uncertainty wavepacket
\begin{equation}
\chi(X) = \left(\dfrac{1}{2\pi \sigma^2_{X}}\right)^{1/4} \exp{ \left[ -\frac{(X-X_0)^2}{4\sigma^2_{X}} \right]}
\exp {\left[ i K_0 X \right]}, \label{gauss_in_st}
\end{equation}
centered in the initial CM position $X_0$ with width $\sigma_{X}$, and propagating with an average CM momentum
\begin{eqnarray}
K_0 \equiv \sqrt{\frac{2M}{\hbar^2} E_{CM}},
\end{eqnarray}
where $E_{CM}$ is the most probable CM kinetic energy of the IX with total mass $M$.

\subsection{Analysis of the time-dependent dynamics}

In order to analyze quantitatively the time-dependent dynamics of the IX, we compute the quantum probability
\begin{equation}
P_{\Omega}(t) \equiv \iint_{\Omega} dx dX |\Psi(X,x,t)|^2. \label{prob_int}
\end{equation}
By choosing the integration domain, $\Omega$, as a suitable sub-domain of the simulation region (details are given in the following sections), we can quantify the transmission, reflection, and dissociation probabilities.

Furthermore, we analyze excitations of internal degrees of freedom of the IX, which may take place due to the coupling between the internal degrees of freedom and the scattering potential, projecting the wavepacket on the eigenfunctions $\phi_n(x)$ of the free Hamiltonian $\hat{H}_0$. Therefore, at any desired time $t$ we compute\footnote{The $\phi_n(x)$ can be chosen to be \textit{real} valued functions.}
\begin{eqnarray}
c_n(X,t) \equiv \int_\Omega \phi_n(x) \Psi(X,x,t) d x \label{coeff_n},
\end{eqnarray}
where $\Omega$ is, as in Eq.~(\ref{prob_int}), a suitable sub-domain. This quantity is further squared to eliminate complex phases and averaged over the CM coordinate, i.e.
\begin{eqnarray}
\left\langle |c_n(X,t)|^2 \right\rangle_{X} \equiv \int |c_n(X,t)|^2 d X .
\end{eqnarray}
Moreover, we normalize the obtained coefficients, namely
\begin{eqnarray}\label{eq:projection}
\langle |c_n|^2 \rangle (t) \equiv \dfrac{\left\langle |c_n(X,t)|^2 \right\rangle_{X}}{\sum_m \left\langle |c_m(X,t)|^2 \right\rangle_{X}}.
\end{eqnarray}
\section{Space- and time-dependent dynamics}
\label{sec:TDdynamics}

We performed extensive numerical investigations of the wavepacket propagation in a broad range of the potentials intensities, $U^e_0$ and $U^h_0$, considering both the attractive and the repulsive case. All simulations have been performed with typical GaAs parameters, as this is the material of choice for IX generation and propagation in CQW structures: $m_e = 0.067 \, m_0$, $m_h = 0.45 \, m_0$, where $m_0$ is the free electron mass, and $\epsilon_r = 12.9$, corresponding to an effective Bohr radius $a_B =11.7\,\mbox{nm}$. All simulations are performed with $d=17\,\mbox{nm}$.\cite{Schinner_PRL13} The eigenvalues of the IX free 1D Hamiltonian $H_r$ obtained by straightforward numerical diagonalization are reported in Tab.~\ref{tab:energies}.

The initial wavepacket has been chosen as described in Sec.~\ref{init_state} with a most probable CM kinetic energy of $E_{CM}=0.5$~meV. Typical grid point densities of 2.5 and 0.75 points/nm have been used for the $x$ and $X$ coordinate, respectively. Sharp potential profiles have been used with $a=2$~nm. Propagation has been usually performed up to 40 picoseconds, with a time step $\Delta_t = 20\, \mbox{fs}$, with $(x,X) \in [-0.6 ,0.6 ]\mathrm{\mu m} \times [-1.5,1.7]\mathrm{\mu m}$.

\begin{table}
\begin{tabular}{cc}
 $n$ & $E_{r,n} (\mbox{meV})$ \\
  \hline
  0 & -4.63 \\
  1 & -2.18 \\
  2 & -1.26 \\
  3 & -0.80 \\
  4 & -0.55 \\
  5 & -0.40 \\
  \hline
\end{tabular}
\caption{Lowest six out of the eight negative eigenenergies of the free IX relative-coordinates Hamiltonian Eq.~(\ref{rel_ham}) for the parameters used in the simulations (see text).}
\label{tab:energies}
\end{table}

\subsection{Potential Step}

In this subsection we analyze the scattering against potential steps or downhills, as described by Eq.~(\ref{pot_step}). It turns out that there are four typical phenomenologies, which we discuss below.

\subsubsection{Total reflection}

\begin{figure}[!h]
\begin{center}
\includegraphics[scale=.18]{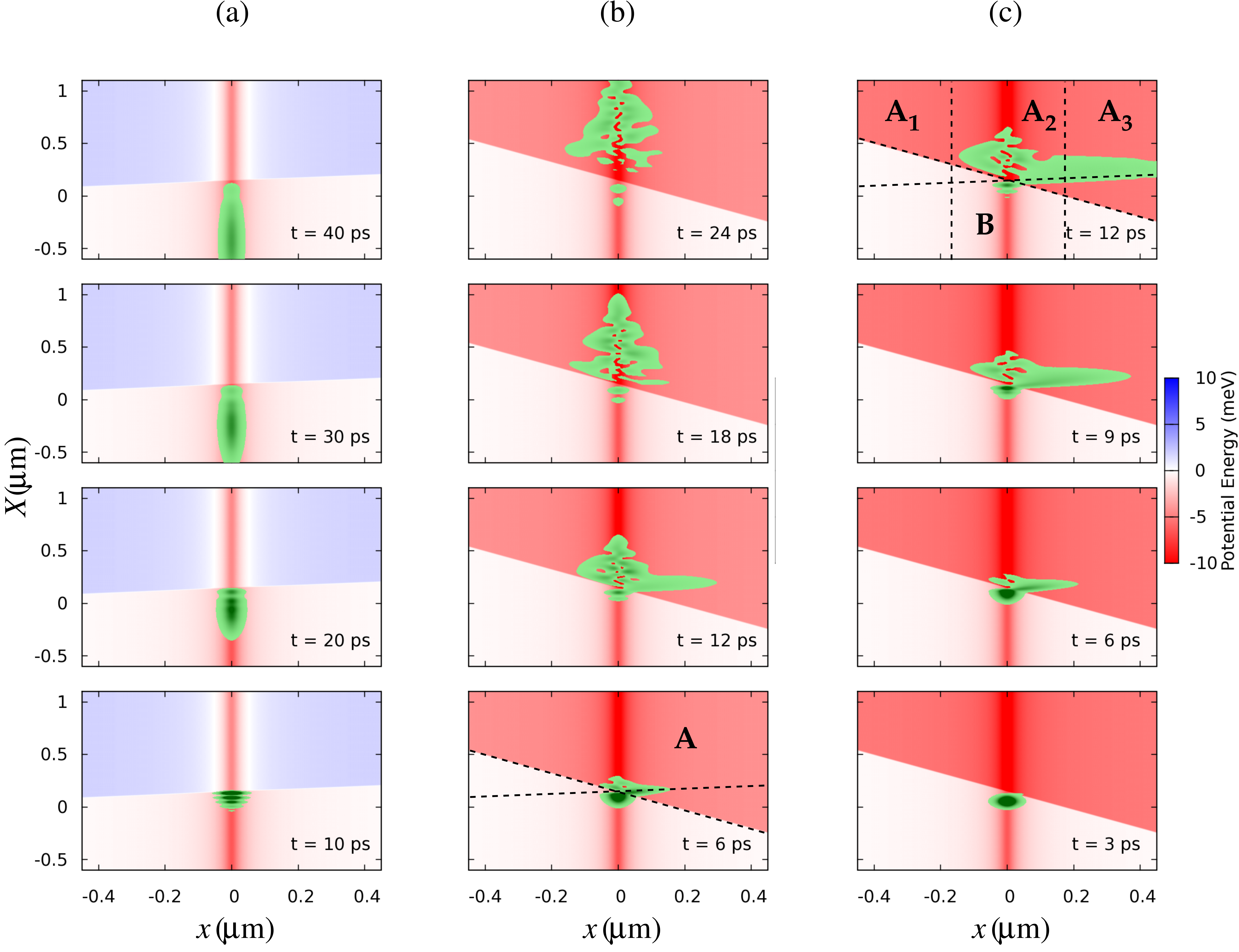}
\end{center}
\caption{Evolution of a IX wavepacket scattering against a potential step/downhill set at $b=150$~nm. The origin of the CM coordinate is set at the center of the $t=0$ wavepacket. In all cases, only one particle is accelerated: (a) $U_0^e = 0.0$~meV, $U_0^h = +2.0$~meV. (b) $U_0^e = -4.0$ meV, $U_0^h = 0.0$ meV. (c) $U_0^e=-5.0$~meV, $U_0^h=0.0$~meV. In (b) and (c) labels mark the integration domains $\Omega$ used in Eqs.~(\ref{coeff_n}),(\ref{prob_int}) (see text), limited between dashed lines corresponding to $x_e=x_h=b$ (slanted lines) and $x=\pm 15 a_B$ (vertical lines). A short animation of the continuous time evolution in (a) is included in the Supplemental Material.\cite{supmat}}
\label{fig:step}
\end{figure}

In Fig.~\ref{fig:step}(a) we show a typical evolution with $U_0^h>E_{CM}$ and $U_0^e = 0$. This case resembles the textbook example of a particle reflected by a potential step, with interference fringes forming between incoming and reflected parts of the wavepacket. After the process is completed, the square modulus of the wavepacket takes the original gaussian shape, spread in the CM coordinate according to the free single-particle propagation, with the standard deviation which can be obtained analytically to be $\sigma_X(t)=\sigma_X(t=0)\sqrt{1+\{t \hbar/[2M\sigma_X(t=0)]\}^2}$.\cite{Mark_EJP97}
This is the general behavior when $U_0^e+U_0^h \gtrsim E_{CM}$. Although it is in principle possible to transfer kinetic energy from the CM to the relative motion and excite IX internal levels, for the present parameters this does not happen, since the CM kinetic energy $E_{CM}$ is substantially smaller than the lowest excitation gap $E_{r,1}-E_{r,0}= 2.46 \,\mbox{meV}$ of the relative coordinate.

\subsubsection{Transmission}
In Fig.~\ref{fig:step}(b) we show the complementary situation of an electron undergoing the acceleration by a downhill, while the hole does not experience any potential. For this particular case, almost all the wavepacket is transmitted, as one would expect for a single particle scattering. However, here the transmission process is more complex. At intermediate times, for $\approx 12$ ps, part of the wavepacket is strongly displaced from $x =0$, that is the electron moves far from the hole, while the CM remains almost still (due to the larger hole mass, the CM position corresponds nearly to the hole coordinate). In this transient, the electron is accelerated, while the hole is not. At a later time, due to Coulomb interaction, the electron drags the hole along. In a classical picture, a wide oscillatory motion of the two particles is activated some picoseconds after the collision. In quantum terms, this corresponds to the excitation of higher quantum levels of the excitonic complex, which can be seen from the complex shape of the transmitted wavepacket in Fig.~\ref{fig:step}(b).

\begin{figure}[!h]
\begin{center}
\includegraphics[scale=.7]{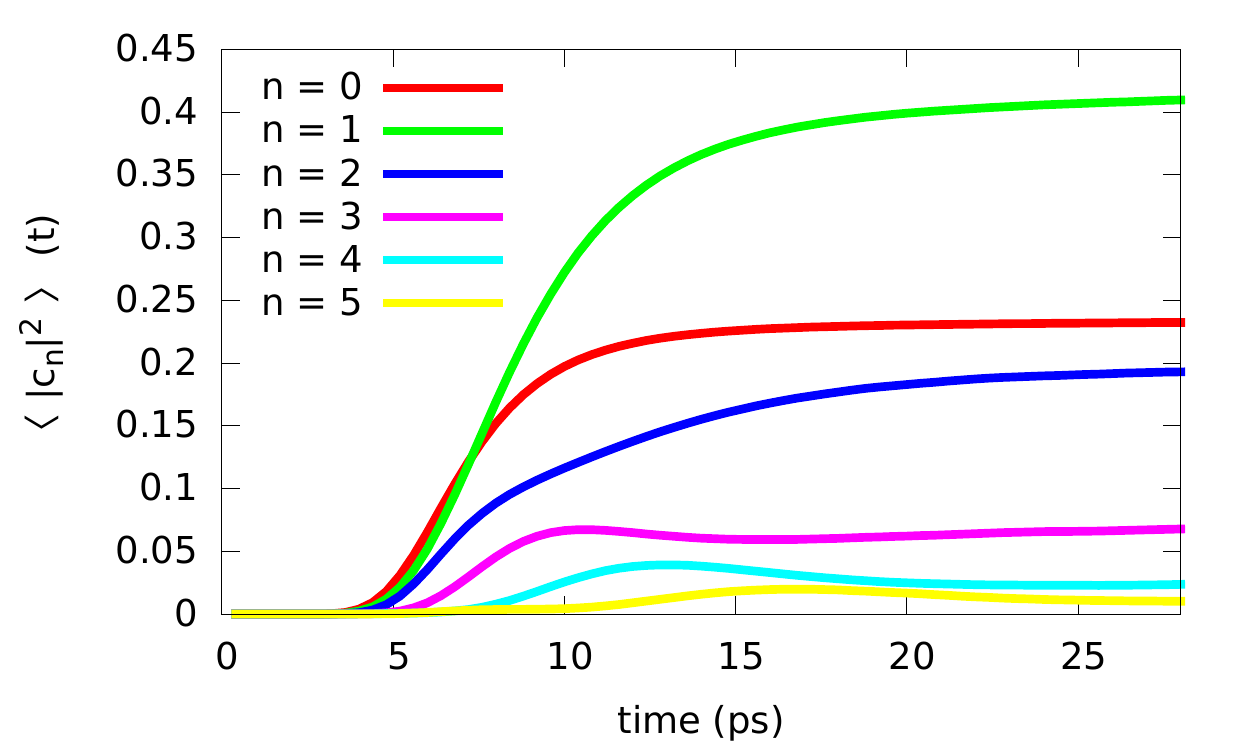}
\end{center}
\caption{Evolution of $\langle |c_n|^2 \rangle (t)$ [see Eq.~(\ref{eq:projection})], with $\Omega \equiv\textbf{A}$ [see Fig.~\ref{fig:step}(b)].  \label{eigR013}}
\end{figure}

To analyze quantitatively the evolution, in Fig.~\ref{eigR013} we show the projection coefficients, $\left\langle |c_n|^2 \right\rangle(t)$. The average in the $X$ coordinate is performed over the transmission region \textbf{A} indicated in Fig.~\ref{fig:step}(b). The scattering starts at $t \approx 3$~ps. Higher internal levels are increasingly excited and the scattering process is completed after $\approx 20 \, \mbox{ps}$. The largest part of the wavepacket ($\approx 40\%$) is in the first excited ($n=1$) state, while another $\approx 40\%$ is almost equally contributed by states $n=0\,\mbox{and}\,2$. All contributions sum up to $\approx 96\%$ of the wavepacket, which is the transmission probability. Note that different excited states are activated at different times and the most advanced part is of the $n=0$ character.

\subsubsection{Dissociation}
Similarly to excitation to higher IX states, dissociation into an unbound electron-hole pair is possible if
\begin{eqnarray}
\left(E_{r,0} + E_{CM}\right) - (U_0^e + U_0^h) > 0,
\end{eqnarray}
that is, an amount of energy exceeding the IX binding energy $E_B = -E_{r,0}$ can be transferred to the relative degree of freedom. In our simulations, since $E_{CM} \ll E_B$ we can observe dissociation only with $-(U_0^e + U_0^h) \lesssim 4.1$~meV. In our numerical evolution, dissociation occurs when the wavepacket asymptotically departs from $x=0$ after scattering. In order to quantify this phenomenon, we compute $P_{\Omega \equiv \mbox{\textbf{\scriptsize A}}_3}(t)$, i.e., we integrate the probability density in a region far from $x=0$ [see Fig.~\ref{fig:step}(c)], and we define the \textit{dissociation probability}
\begin{eqnarray}
P_{diss} \equiv \iint_{\mathrm{\textbf{A}}_3} dx dX |\Psi(X,x,t>t_\infty)|^2.
\end{eqnarray}
where $t_{\infty}$ is a sufficiently large time.

\begin{figure}[!h]
\begin{center}
\includegraphics[scale=0.7]{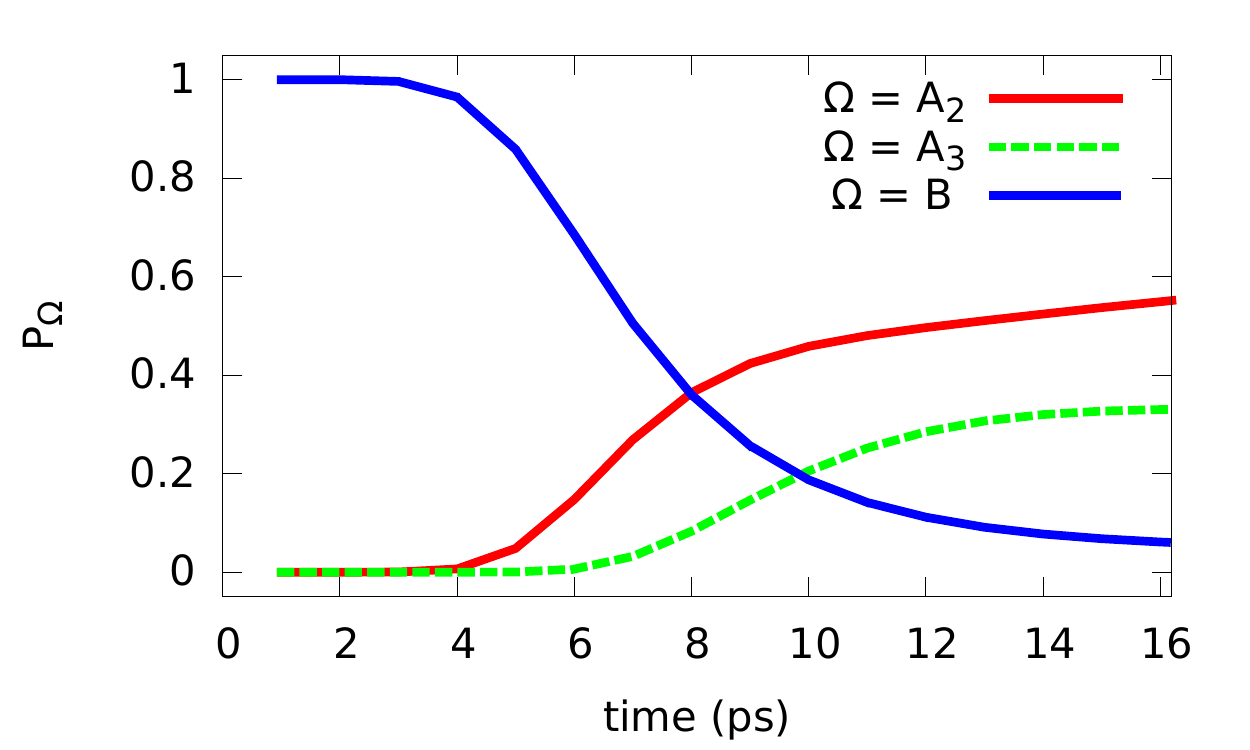}
\end{center}
\caption{Evolution of $P_\Omega (t)$ [see Eq.~(\ref{prob_int})], with $\Omega$ coinciding with regions \textbf{A}$_2$, \textbf{A}$_3$, or the central region \textbf{B} indicated in Fig.~\ref{fig:step}(c). The probability over the region \textbf{A}$_1$ is vanishing and it is not shown. \label{int10_R001} }
\end{figure}

In Fig.~\ref{fig:step}(c) a part of the wavepacket is pulled off to the right.
Here $P_{diss}\approx 30$\%, as shown in Fig.~(\ref{int10_R001}).
Note that in this particular simulation $U_0^h=0$, while $U_0^e > E_B$.
Therefore, the electron acquires a high amount of energy from the downhill potential, this exciting the internal dynamics of the IX which eventually dissociates. This is similar (with role of the electron and the hole inverted) to Fig.~\ref{fig:step}(b), where, however, $U_0^h<E_B$ and the IX is excited but does not dissociate.

\subsubsection{Dwelling nearby $X=b$}
\begin{figure}[!h]
\begin{center}
\includegraphics[scale=.18]{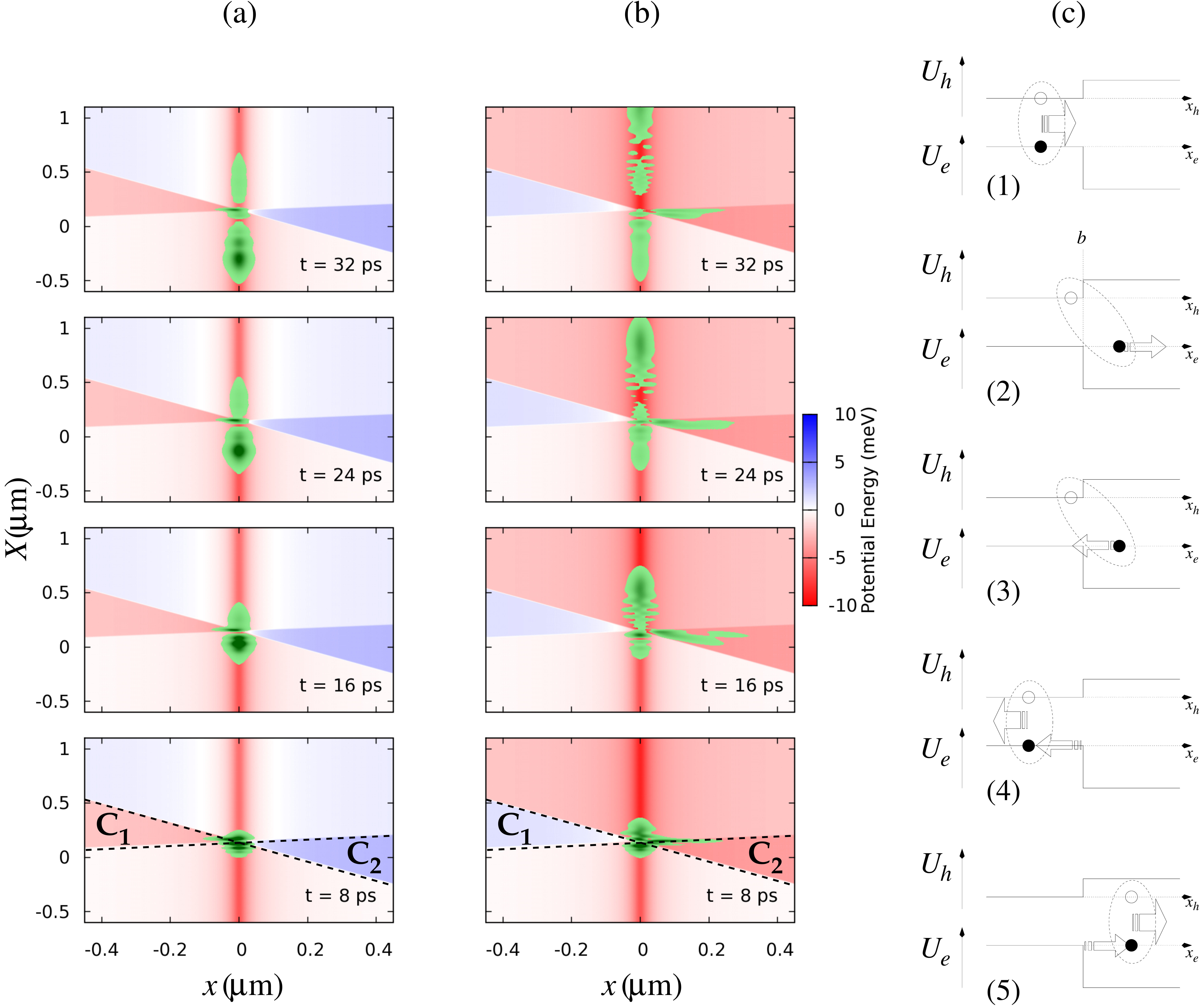}
\end{center}
\caption{As in Fig.~\ref{fig:step} with (a) $U_0^e=+3.0$~meV, $U_0^h=-2.0$~meV.(b) $U_0^e=-3.5$~meV, $U_0^h=1.5$~meV. (c) Semiclassical interpretation of the scattering process shown in (b). Dashed lines represent equations $x_e=b$ and $x_h=b$. \label{pause}}
\end{figure}

As a last interesting phenomenology arising from the two-body dynamics, we analyze cases in which part of the exciton wavepacket dwells in the proximity of $X=b$, i.e., near the step, for a long time interval. This occurs when both the electron and the hole experience a strong potential, but with opposite sign, a step for the electron and a downhill for the hole or \textit{vice versa}. Figures~\ref{pause}(a),(b) shows the two cases. Dwelling may occur in region \textbf{C}$_1$ [Fig.~\ref{pause}(a)] or \textbf{C}$_2$ [Fig.~\ref{pause}(b)], depending on the electron/hole potentials. A semiclassical interpretation of the process is given in Fig.~\ref{pause}(c) with reference to the case shown in Fig.~\ref{pause}(b). As the IX wavepacket hits the scattering potential (1), the electron is accelerated along the downhill, while the hole does not overcome the repulsive step despite the Coulomb attraction from the partner particle (2). Then, the electron, which is lighter, is pulled backwards by the Coulomb interaction (3). Part of its wave function is transmitted by the now-repulsive step, joins the hole at a small kinetic energy, and the whole IX is reflected backwards (4), while part of the electron wave function is reflected from the step, dragging the hole along, and the exciton as a whole is transmitted forward (5). However, there is a probability that, after reflection, the electrons does not pull the hole far from the barrier. In this case, the electron perform one more oscillation, and the process starts over again. Therefore, in such a process reflection and transmission occur as a periodic \emph{pumping} of smaller wavepackets. The description of the process in Fig.~\ref{pause}(a) is analogous with the roles of the electron and the hole inverted.

\subsection{Potential Barrier}

We now investigate the scattering from a finite range potential, Eq.~(\ref{pot_barr}). A barrier/well for each carrier is present in a region $b_2-b_1=40$~nm wide. Note that, while the width of the barrier/well has an obvious effect on the scattering event, its absolute position is unimportant, as far as the two parameters $b_1$ and $b_2$ are the same for the two particles.
In the present case quantum tunneling becomes possible, as in usual single-particle scattering. On the other hand, in the asymptotic regions $U_0^e = U_0^h = 0$. Therefore, it is not possible to excite internal levels or to have free particles, since, as discussed in the previous section, the CM kinetic energy which could be transferred to the internal degrees of freedom is much lower than the excitation gaps of the IX and, \emph{a fortiori}, its binding energy. However, an oscillatory motion or even the dissociation of the IX may occur if one of the particle is captured by its well, transferring a large energy to the partner particle.

\begin{figure}[!h]
\begin{center}
\includegraphics[scale=.18]{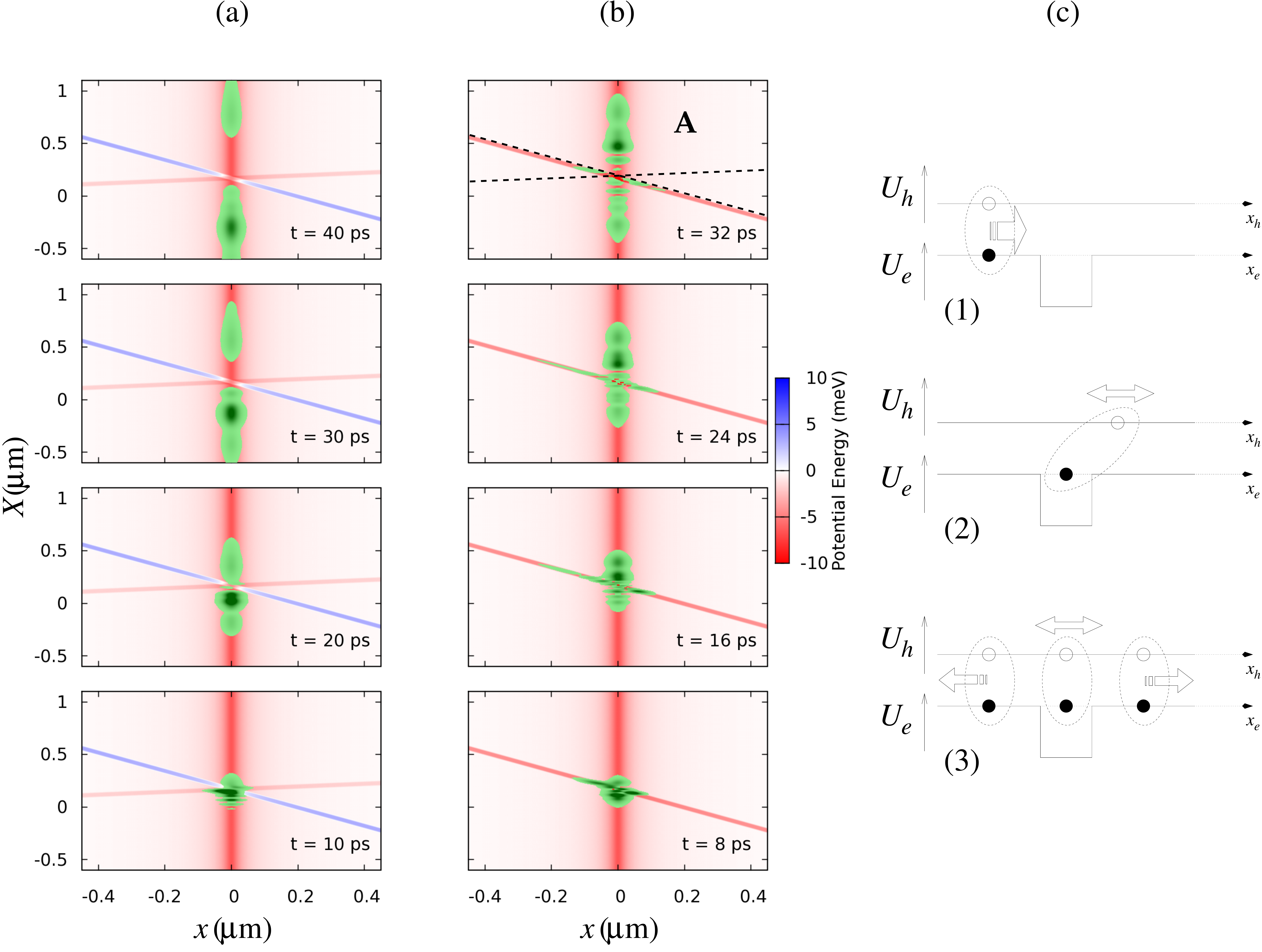}
\end{center}
\caption{Scattering through potentials barriers/wells with $b_1 =150\,\mbox{nm}, b_2 =190\,\mbox{nm}$. (a) $U_0^e=3.5$~meV, $U_0^h=-1.5$~meV. (b) $U_0^e=-4.0$~meV, $U_0^h=0.0$~meV. (c) Sketch of the periodic emission of the IX wavepacket. Short animations of the continuous time evolution in (a) and (b) are included in the Supplemental Material.\cite{supmat}
\label{transm_barr}}
\end{figure}

\begin{figure}
\begin{center}
\includegraphics[scale=.6]{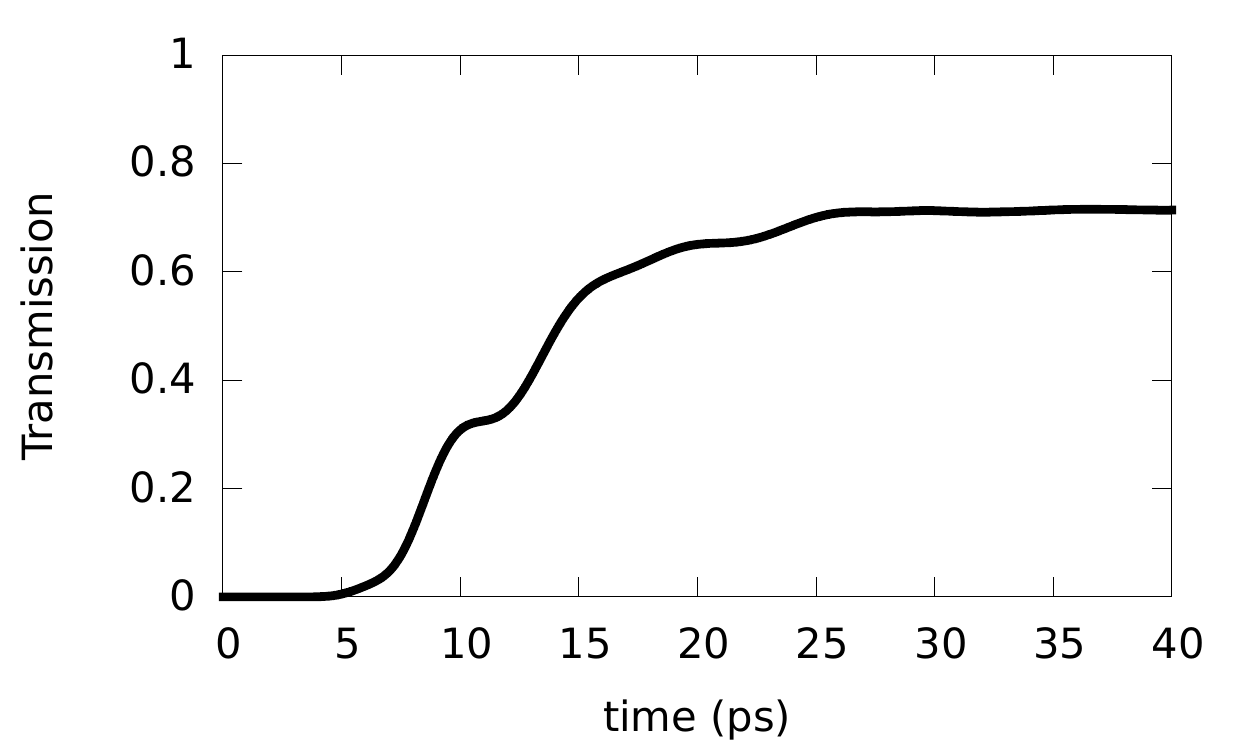}
\end{center}
\caption{Transmission coefficient calculated from Eq.~(\ref{prob_int}) with $\Omega = \textbf{A}$ [see Fig.~\ref{transm_barr}(b)]. \label{transm_vs_time}}
\end{figure}

In Fig.~\ref{transm_barr}(a) a wavepacket hits a high barrier for electrons and a weaker well for the hole and tunnels with $\approx 19$\% probability. Note that the hole dwells into its potential well in the initial stages of the process. Dwelling may also lead to periodic emission of the wave function, as for the step/downhill case, which is shown in Fig.~\ref{transm_barr}(b) and sketched in  Fig.~\ref{transm_barr}(c). When the IX hits the barrier/well potential (1), the electron may be captured by its well, while the hole starts to oscillate around the well position (2). At each oscillation, the wavepacket located in the well region is partially transmitted and partially reflected as a IX bound state (3). These oscillations are  highlighted by the step-like evolution of the transmission coefficient, shown in Fig.~\ref{transm_vs_time}.

\section{Scattering phase diagram}
\label{sec:scattering}

The space- and time-dependent evolutions discussed in the previous section are selected examples in a large set of simulations, to highlight how the scattering process takes place in different regimes. In this section we summarize the outcome of the scattering events at asymptotic times in the full parameter space $U_0^e$ and $U_0^h$.

For each simulation we computed transmission, reflection, and excitation probabilities at $t=40$~ps,\footnote{For selected simulations, when the IX dissociates, we stopped the simulation at shorter times, i.e., when one of the particles was hitting the boundary of the simulation range.} that is, at a sufficiently large time for the scattering process to be completed. These coefficients are used to identify distinct phenomenologies taking place at different values of the potential parameters $U_0^e$ and $U_0^h$. For each class of potentials, either steps/downhills or barriers/wells, we summarize our results in a phase diagram as a function of the potential total intensity $(U_0^e + U_0^h)$ and potential asymmetry $(U_0^e - U_0^h)$ between the two particles, rather then $U_0^e$ and $U_0^h$ separately, as the former quantities are those driving the distinct behaviors. A common energy scale for all simulations is set by the kinetic energy $E_{CM}=0.5\,\mbox{meV}$. In the phase diagrams to be discussed below, it is convenient to report results in terms of the dimensionless ratios $(U_0^e + U_0^h)/E_{CM}$, $(U_0^e - U_0^h)/E_{CM}$. Note that for a rigid exciton, i.e., an exciton with frozen internal degrees of freedom,\cite{Zimmermann_PAC97} the scattering outcome (i.e., transmission probability) scales with these ratios, similarly to the textbook single particle scattering. However, in the present case of a quantum system with an internal degree of freedom, the latter couples to the CM kinetics and is affected by the external potentials. Therefore, here the scaling is only approximate: although still a convenient representation, one should keep in mind that different behaviors might be expected in a given region of the phase diagram if $E_{CM}$ varied. This is particularly true for what concerns the phenomenologies arising from the internal dynamics, i.e. dissociation, excitation to higher levels, pausing/pumping nearby the external potential region.

\subsection{Potential steps/downhills}

\begin{figure}
\begin{center}
\includegraphics[scale=.3]{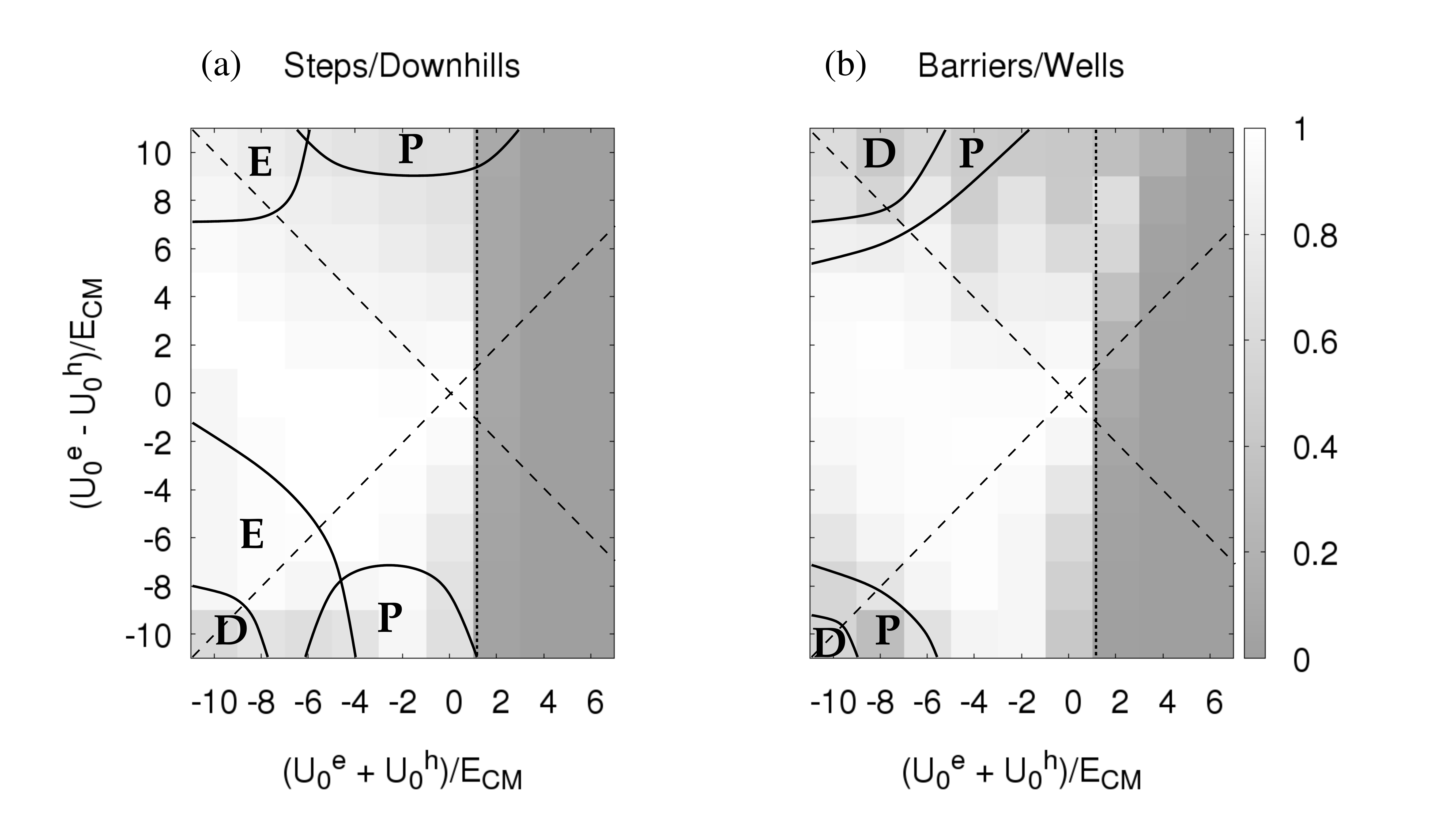}
\end{center}
\caption{Phase diagram for (a) steps/downhills and (b) barriers/wells simulations. The transmission probability is displayed in grayscale. Simulations are performed every 1 meV in $(U_0^e + U_0^h)$ and $(U_0^e - U_0^h)$. The main phenomenologies are highlighted by superimposed letters and solid lines to mark approximate boundaries of different regions.  Regions \textbf{D}: dissociation probability greater than about $5$\%. Regions \textbf{E}: excitation to higher internal levels greater than about $10$\%. Regions \textbf{P}: periodic transmission/reflection of the wave function due to dwelling. The positive (negative) slope dashed line indicates $U_0^{h(e)}=0$. }
\label{PD}
\end{figure}

\subsubsection{Transmission/reflection}
In Fig.~\ref{PD}(a), the transmission probability is reported as a gray scale colormap. Simulations are performed on a coarse grid with steps of 1 meV in $(U_0^e + U_0^h)$ and $(U_0^e - U_0^h)$, i.e., in steps of 2 units in the ratios $(U_0^e \pm U_0^h)/E_{CM}$. This choice of the simulation space is a good trade-off between a sufficiently detailed insight and the computational cost. In Fig.~\ref{PD}(a), each of the 242 coloured squares corresponds to a specific simulation running for about 8 hours (including postprocessing) on a high-end workstation. When the IX wave function leaves the central region of the domain, i.e. the active region where the external potential may cause reflections or trigger excited internal states, the transmission probability is computed by means of Eq.~(\ref{prob_int}). The integration region $\Omega$ is indicated as $A_2$ in Fig.~\ref{fig:step}(c) and corresponds to the probability of finding both the electron and the hole beyond the step ($x_e>b$ and $x_h>b$) and less than 15$a_B$ apart ($x<15a_B$). 

As might be expected, transmission is large on the left part of the diagram, corresponding to electron and hole downhills, and drops on the right hand side, which correspond to repulsive steps for both particles, vanishing at $(U_0^e + U_0^h)/E_{CM} \simeq 1$. Note that if the potential is symmetric for the two particles, i.e., $(U_0^e - U_0^h) \simeq 0$, the transmission rapidly drops from 1 at $(U_0^e + U_0^h)/E_{CM} \simeq 0$ to zero. However, for strongly asymmetric potentials $| (U_0^e - U_0^h)/E_{CM}| \gtrsim 5$, that is in the upper and lower parts of the diagram, the transmission start to decrease at negative values of the total potential $U_0^e + U_0^h$ and evolves smoothly to zero. This is because in these regions, one particle experiences a strongly attractive downhill potential, while the other is subject to a weaker but repulsive potential. Note also that the diagram is slightly asymmetric in the upper and lower parts. This is due to the different masses of electrons and holes, so that similar potential intensities, but with the role of the particles inverted, have a different effect on the transmission.

\subsubsection{Excitation/Dissociation}
Excitations of the relative motion eigenstates $\phi_n(x)$ are favored when both $(U_0^e + U_0^h) < 0$ and $(U_0^e - U_0^h)  <0 $ [lower left corner of Fig.~\ref{PD}(a)], which corresponds to a large downhill potential for the electron and a weaker potential step for the hole. Clearly, the situation may be reversed, with electron and hole potentials inverted, $ (U_0^e - U_0^h)>0$ (top left corner). However, in this case the excitation is typically weaker, so that regions of excitations and dissociation move to stronger (positive) anisotropies in Fig.~\ref{PD}(a).

This asymmetry can be understood from inertial considerations: since the hole is one order of magnitude heavier than the electron, a step/downhill for the hole almost coincides with the potential acting on the CM coordinate [a different way to see this is to note that the potential step for the hole is nearly independent from the relative coordinate, compare Figs.\ref{fig:step}(a) with Figs.\ref{fig:step}(b),(c)], and the IX accelerates as a single, rigid particle. On the other hand, when an electron experiences an equivalent downhill potential, a larger part of the external energy can be transferred to the internal relative dynamics. Dissociation can also take place, when the electron downhill is particularly strong.\footnote{In practice, in the simulations IX dissociation is signalled by excitation of higher and higher internal levels, since an increasing number of bound states is be necessary to describe the free part of the wavepacket as time proceeds.}

\subsection{Potential barriers/wells}

Contrary to the step/downhill case, with a barrier/well potential it is not possible to have excitation of internal levels nor dissociation in the transmitted part of the wavepacket.  As discussed in the previous section, however, for $(U_0^e + U_0^h)/E_{CM} \ll 0$ and $|U_0^{e(h)}|\gg E_{CM}$ dissociation \textit{in the barrier region} takes place if one particle (the electron or the hole, depending whether $U_0^{e}$ or $U_0^{h}$ is negative and intense) is captured into its well, while the other inertially continues its motion.

Due to quantum tunneling, the transmission probability is finite also for $(U_0^e + U_0^h)/E_{CM} \geq 1$. Note that tunneling is substantially anisotropic between the upper part ($U_0^e > 0$ and $U_0^h<0$) and the lower part ($U_0^e<0$ and $U_0^h>0$) of the phase diagram. Again, this is clearly due to the difference between electron and hole effective masses, as for large anisotropies IX does not tunnel as a rigid complex of mass $M$, but rather through a complex two-body process. On the other hand, as one may expect from one-body tunneling, resonant transmission may take place. This is shown in Fig.~\ref{PD}(b) by the oscillatory behavior of the transmission coefficient, particularly in the upper part of the phase diagram.

Finally, note that, at difference with the steps/downhills case where dissociation and excitation take place more easily for negative anisotropies (lower left part of the phase diagram) than for positive anisotropies (upper left part), here the opposite is true, since these phenomena happen by the trapping mechanism, which is much more favorable for holes.

\section{Conclusions}
\label{sec:conclusions}

We have discussed the unitary space- and time-dependent quantum dynamics of a single IX in a CQW system scattering against simple potential profiles. Our analysis suggests, in particular, that exciton scattering in typical devices is a genuine two-body process, substantially different and more complex than the single-particle scattering of a rigid exciton. Our results are based on a minimal 1D model. Although strictly speaking this analysis applies to 1D channels (e.g., electrostatically defined coupled quantum wires), it allows to capture the main phenomenologies to be expected also in planar CQWs. Next we discuss the connection to realistic systems and perspective experiments.

For the prototypical potential profiles considered here (but in a large range of parameters) IX scattering times turn out to be in the order of tens of ps. This is smaller than the lifetime of IXs in CQWs and also their coherence time in quantum wells, \cite{Enderlin_PhysRevB.80.085301} and supports the possibility to engineer single-IX phase coherent excitronic devices.

Having used typical material parameters and potential landscapes for GaAs-based structures, we expect that the complex phenomenologies found in our simulations could take place in excitronic devices and should be taken into account in device engineering. Moreover, the complex time-dependent IX dynamics exposed by our results could be directly probed, for example, by collecting time-resolved micro-photoluminescence maps \cite{High_Science11072008,Gartner_APL06} \emph{during} the scattering process, exploiting the fact that optical recombination of IXs can be triggered at a desired time by switching off the vertical field $F_z$.\cite{Lozovik_SSC01,Winbow_JAP08} Luminescence maps
could be calculated\cite{Feldmann_PRL87,Andreani1991641,Butov_NatureB02} from $|\Psi(X,x=0)|^2$, as provided, e.g., by our simulative method. Although we did not attempt an explicit calculation of optical signatures to be directly compared to experiments, since our model is strictly 1D, for purpose of illustration we provide in the Supplemental Material\cite{supmat} the real-space and K-space evolution of $\Psi(X,x=0)$.
Note that in this respect single-IX devices could offer a unique possibility to probe the quantum two-body scattering dynamics by optical means. In realistic 2D devices, scattering times could be larger than those found in our simulations, due to the larger phase space. Time-resolution can thus be less demanding.

Potential profiles as those employed in the present study are usually induced by electrostatic gates. On the one hand, typical top gates induce an electrostatic potential in the underlying 2D system with opposite sign for the two particles. Due to layer separation $d$, however, the potential generated by a top gate is different in magnitude for the two layers. For example, for a device of total thickness ~1 $\mu$m with ~1 V applied to top gates the resulting voltage difference between the two CQWs 10 nm apart is of the order of ~10 mV. Therefore, a common situation in excitronic devices is with large potential anisotropies, corresponding to the upper and lower sectors between the bisecting lines in Fig.~\ref{PD}.

On the other hand, it is also possible to gate separately the two layers of the heterostructure and engineer potential landscapes of the same sign and similar intensities for the two particles \cite{Eisenstein_APL91}. Potential profiles of the same sign for the two particles with few meV anisotropies correspond to left and right regions between the bisecting lines in Figs.~\ref{PD}. Scattering potentials of the same sign may also arise from monolayer fluctuations of the QW thickness which induce modulations of the lateral confinement energy in the meV range  i.e, a steps/downhills potential profiles along the planes of the QWs.

\section{Acknowledgments}

We thank M. Rontani for a critical reading of the manuscript and L. Sorba for useful discussions.
We acknowledge partial financial support from EU-FP7 Initial Training Network INDEX and University of Modena and Reggio Emilia, Grant ``Nano- and emerging materials and systems for sustainable technologies''.

\appendix

\section{The two-particle Split-Step method}
\label{appendix_a}

In order to study the time-dependent scattering dynamics of IXs, we use a numerically exact solution of the Time Dependent Schr\"odinger Equation (TDSE) with the Hamiltonian given by Eq.~(\ref{ham_ecc_lib}) based on the Split Step Fourier method.

The time propagation of a generic quantum state $\Psi(\mathbf{x}, t)$, between two close time instants, $t$ and $t + \Delta_t$, can be expressed by means of the evolution operator, $\hat{\mathcal{U}}(t+\Delta_t,t)$, as:
\begin{equation}
\Psi(\mathbf{x}, t +\Delta_t) = \hat{\mathcal{U}}(t+\Delta_t,t) \Psi(\mathbf{x}, t),
\end{equation}
where:
\begin{eqnarray}
\begin{split}
&\hat{\mathcal{U}}(t + \Delta_t, t) = \\
&= \mathcal{T} \exp \left\lbrace -\frac{i}{\hbar} \int_t^{t+\Delta_t} \left( \hat{T} + \hat{U}(\tau) \right) d\tau \right\rbrace \approx\\
& \approx \mathcal{T} \exp \left\lbrace -\frac{i}{\hbar} \left( \hat{T} \Delta_t + \hat{U} \Big(t + \frac{\Delta_t}{2} \Big) \Delta_t \right) \right\rbrace. \\ \label{ss1}
\end{split}
\end{eqnarray}
Here $\mathcal{T}\exp\{\}$ is the so-called \textit{time ordered exponential operator}. In the last line the `rectangle integration method' has been used for integrating the potential energy.
The last line in Eq.~(\ref{ss1}) is exact for time independent potentials.

By using the Suzuki-Trotter formula\cite{Suzuki_PJAB93} in order to expand the exponential of the sum of operators in Eq.~(\ref{ss1}) as a product of exponential operators, we obtain
\begin{eqnarray}
\begin{split}
&\hat{\mathcal{U}}(t + \Delta_t, t) =\\
&= e^{ -\frac{i}{\hbar} \hat{U} (t + \frac{\Delta_t}{2} ) \frac{\Delta_t}{2} }
          e^{ -\frac{i}{\hbar} \hat{T}  \Delta_t } e^{ -\frac{i}{\hbar} \hat{U} (t + \frac{\Delta_t}{2} ) \frac{\Delta_t}{2}} + \mathcal{O}(\Delta_t^3).\label{ss2}
\end{split}
\end{eqnarray}

We now exploit the fact that the operator $ \hat{T} $ is diagonal in Fourier space, while the operator $\hat{U} $ is diagonal in direct space.
The basic idea behind the \textit{Split Step Fourier} (SSF) method is to exploit the evolution operator factorization, Eq.~(\ref{ss2}), in order to act on the state function in the specific representation in which the operator is diagonal. At each time step the SSF algorithm performs the following operations:
\begin{enumerate}
\item Multiplication of the wave function in coordinates representation by the rightmost potential term, for half time step:
\begin{equation}
\Psi(\mathbf{x},t) \longrightarrow \Psi '(\mathbf{x}) = e^{-\frac{i}{\hbar} \hat{U}\left(\mathbf{x}, t + \frac{\Delta_t}{2}\right) \frac{\Delta_t}{2}} \Psi(\mathbf{x}, t) .
\label{eqA_a4}
\end{equation}

\item Computation of the Fourier transformation of the state $ \Psi '(\mathbf{x}) $:
\begin{equation}
\tilde{\Psi}'(\mathbf{p}) = \int {d\mathbf{x} e^{-i\mathbf{p}\cdot \mathbf{x}} \Psi '(\mathbf{x}) }.
\end{equation}

\item Multiplication of the wave function in momentum space by the kinetic energy term for a full time step, $\Delta_t$:			
\begin{equation}
\tilde{\Psi}'(\mathbf{p}) \longrightarrow \tilde{\Psi}''(\mathbf{p}) = e^{-\frac{i}{\hbar} \sum_i \frac{p_i^2}{2 m_i} \Delta_t} \tilde{\Psi}'(\mathbf{p}).
\label{eqA_a6}
\end{equation}

\item Calculation of the inverse Fourier transform of the wave function:
\begin{equation}
\Psi ''(\mathbf{x}) = \frac{1}{(2\pi)^{2}}\int {d\mathbf{p} e^{+i\mathbf{p}\cdot \mathbf{x}} \tilde{\Psi}''(\mathbf{p}) }.
\end{equation}

\item Multiplication of the wave function by the leftmost potential term for another half time step:
\begin{equation}
\Psi ''(\mathbf{x}) \longrightarrow \Psi(\mathbf{x}, t + \Delta_t) = e^{-\frac{i}{\hbar} \hat{U}\left(\mathbf{x}, t + \frac{\Delta_t}{2}\right) \frac{\Delta_t}{2}} \Psi''(\mathbf{x}).
\label{eqA_a8}
\end{equation}

\end{enumerate}
The SSF is a unitary method which preserves the norm of the wave function.\cite{Leforestier_JCP91}
We underline that in Eq.~(\ref{ss2}) operators $\hat{T}$ and $\hat{U}$ are represented in their respective diagonal forms using the appropriate basis. Therefore, even though a spatial discretization of the domain (as described in Sec.~\ref{sec:TDdynamics}) is used in order to
represent the wavepackets and the potential $U$, no finite difference approach is
needed to represent the kinetic operator, which instead is represented in the reciprocal grid implied by the discrete Fourier transform algorithm.

The SSF method is commonly used to propagate a wavepacket representing one particle in a $n$-dimensional space, e.g. with $\mathbf{x}\equiv (x, y)$ for $n=2$. However, since it is an exact solution of the Schr\"odinger equation, there is no limitation as to the interpretation of the set of coordinates $\mathbf{x}$. Here, we implement the method by propagating the two-body wave function in the $\mathbf{x}\equiv (X,x)$ coordinates. Note that the momentum vector $\mathbf{p}\equiv(P,p)$ has also components corresponding to the CM and relative momenta, respectively.

Unlike the isotropic mass one-body case, the two coordinates are governed by different terms in the Hamiltonian, with very different effective masses, resulting in different frequency ranges for $P$ and $p$. While typical energies for the relative motion are in the few meV range, the typical kinetic energy for the CM is in the fraction of an meV. It is this decoupling which makes particularly convenient to work in the CM and relative coordinates.

\section{Wavepacket in K space}
\label{appendix_b}
\begin{figure}[!h]
\begin{center}
\includegraphics[width=0.9\textwidth, angle=0]{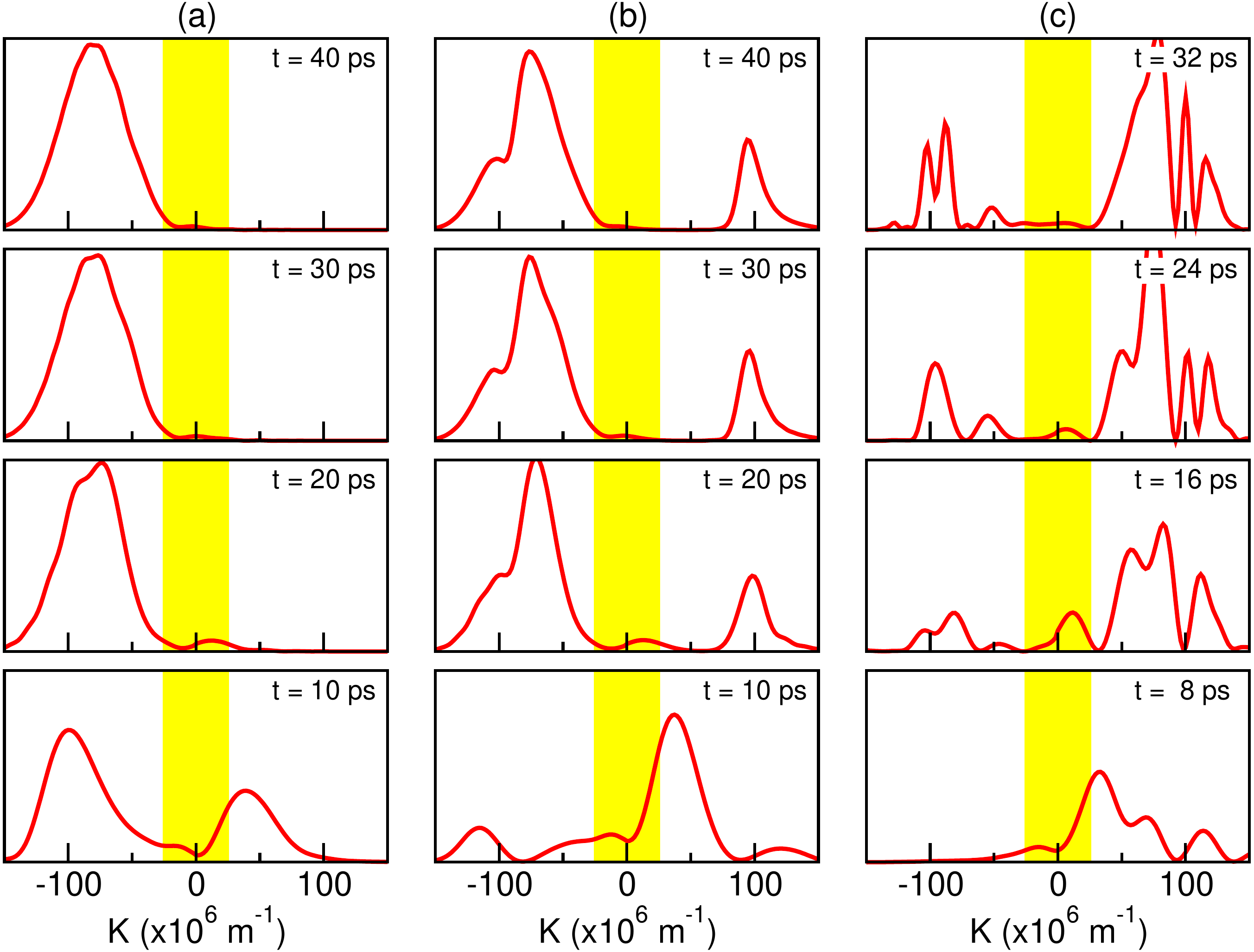}
\end{center}
\caption{$K$-space representation of the wave function at four different times
for the same IX scatterings shown in Figs.~\ref{fig:step}(a), \ref{transm_barr}(a) and \ref{transm_barr}(b), in columns (a), (b) and (c), respectively.}
\label{fig_appendix}
\end{figure}

In Fig.~\ref{fig_appendix} we show the $K$-space representation of $\Psi(X,x=0)$ for the three cases (and the same selected times) presented in Figs.~3(a), 7(a) and 7(b). This quantity is directly related to the luminescence intensity\cite{Feldmann_PRL87,Andreani1991641,Butov_NatureB02} which is $\propto |\Psi(X,x=0)|^2$. The highlighted region around the origin satisfies
\begin{equation}
K < \frac{n}{c\hbar} E_{gap} \ ,
\end{equation}
which represents the $K$-space radiative cone.\cite{Butov_NatureA02}
Here, $n=\sqrt{\epsilon_r}$ is the refractive index of GaAs, $c$ the speed of light, $\hbar$ the reduced Planck constant and $E_{gap}= 1.424$~eV the GaAs energy gap.

The videos \texttt{movieK\_fig3a.mpg}, \texttt{movieK\_fig7a.mpg}
and \texttt{movieK\_fig7b.mpg} of the Supplemental Material\cite{supmat}
show the time evolution of $|\Psi(X,0)|^2$
(top panel) and its $K$-space transform (bottom panel) for the three cases of
Supplementary Fig.~1. The $K$ components that comply with Eq.~(1)
are also highlighted. The evolution shows that such optically-active components
are substantially modified during the scattering of the IX.

Moreover, the videos \texttt{movieXx\_fig3a.mpg}, \texttt{movieXx\_fig7a.mpg}
and \texttt{movieXx\_fig7b.mpg} of the Supplemental Material\cite{supmat},
show the continuous-time evolution of
$|\Psi(X,x)|^2$ for the three scattering events also shown in 
Figs.~\ref{fig:step}(a), \ref{transm_barr}(a) and \ref{transm_barr}(b).

\bibliography{grasselliBIB}

\end{document}